\documentstyle[sprocl,epsf]{article}

\def\Journal#1#2#3#4{{#1} {\bf #2}, #3 (#4)}

\def\NPB{{\em Nucl.Phys.} B}
\def\PLB{{\em Phys.Lett.}  B}
\def\PRL{\em Phys.Rev.Lett.}

\def\ERR{ERR-ibid.}

\def\be{\begin{equation}}
\def\ee{\end{equation}}
\def\bea{\begin{eqnarray}}
\def\eea{\end{eqnarray}}

\def\Title#1{\begin{center} {\Large #1 } \end{center}}
\def\Author#1{\begin{center}{ \sc #1} \end{center}}
\def\Address#1{\begin{center}{ \it #1} \end{center}}
\def\andauth{\begin{center}{and} \end{center}}
\def\submit#1{\begin{center} #1 \end{center}}
\def\doeack{\footnote{Work supported by the Department of Energy,
                     contract DE--AC03--76SF00515.}}
\def\esterack{\footnote{Work supported by the Spanish MEC and FAME under
                      project AEN97--1678.}}
\def\SLAC{Stanford Linear Accelerator Center\\
    Stanford University, Stanford, California 94309 USA}
\newcommand\pubblock{\rightline{\begin{tabular}{l} FTUAM-99-23\\
         SLAC-PUB-8251\\ September 1999 \end{tabular}}}
\newenvironment{Abstract}{\begin{quotation} \begin{center}
                       ABSTRACT
     \end{center}\bigskip  }{\end{quotation}}

\begin{document}
\begin{titlepage}
\pubblock

\vfill
\Title{Probing Strong Electroweak Symmetry Breaking in $W^+W^- 
         \to t \bar t$}
\vfill
\Author{Ester Ruiz Morales\esterack}
\Address{Dept. F\'\i sica Te\'orica, Universidad Aut\'onoma de Madrid\\
     Cantoblanco, 28049 Madrid, SPAIN}
\medskip
\andauth
\medskip
\Author{Michael E. Peskin\doeack}
\Address{\SLAC}
\vfill
\begin{Abstract}
We study the process $W^+W^-  \to t \bar t$ in several models
 of strong interaction electroweak symmetry breaking.
  We calculate the signals that can be expected by observing the
  reaction $e^+ e^-\!\to \nu \bar \nu t \bar t$ at an $e^+ e^-$ linear
  collider with 1.5 TeV center of mass energy. We also discuss how the
  lowest-lying resonances predicted by these models could be
  identified using top polarization observables.
\end{Abstract}
\medskip
\submit{presented at the International Workshop on Linear Colliders\\
          Sitges, Barcelona, Spain, 28 April -- 5 May 1999}

\vfill
\end{titlepage}
\def\thefootnote{\fnsymbol{footnote}}
\setcounter{footnote}{0}

\hbox to \hsize{\null}
\newpage
\setcounter{page}{1}

\title{PROBING STRONG ELECTROWEAK 
SYMMETRY BREAKING IN $W^+ W^-\!\to t \bar t \;\;$} 

\author{ E. RUIZ MORALES }

\address{Dept. F\'\i sica Te\'orica, Universidad Aut\'onoma de Madrid,\\
Cantoblanco, 28049 Madrid, Spain}

\author{ M. E. PESKIN }

\address{Stanford Linear Accelerator Center, Stanford University,\\
Stanford, California 94309, USA}


\maketitle
\abstracts{We study the process $W^+ W^-\!\to t \bar t$ in
  several models of strong interaction electroweak symmetry breaking.
  We calculate the signals that can be expected by observing the
  reaction $e^+ e^-\!\to \nu \bar \nu t \bar t$ at an $e^+ e^-$ linear
  collider with 1.5 TeV center of mass energy. We also discuss how the
  lowest-lying resonances predicted by these models could be
  identified using top polarization observables.}
  
\section{The reaction $W^+ W^-\!\to t \bar t$ in strong-interaction 
Higgs sectors}
\label{sec:1}

Vector boson scattering experiments form the core of the program to 
probe the Higgs mechanism in models with strong interaction electroweak 
symmetry breaking at TeV energies. It was pointed out by 
Barklow~\cite{SMR} that the related process $W^+ W^-\!\to t\bar t$
could also be used to probe how the Higgs sector couples to fermions.
Although QCD backgrounds make this process very difficult to observe
at hadron colliders, he showed that, at an $e^+ e^-$
linear collider (LC) with $\sqrt s = 1.5$ TeV, the signal of a
Standard Model (SM) Higgs sector could be established with
good statistical significance. We have studied the sensitivity of the
reaction $W^+ W^-\!\to t \bar t$ to more general models of strong
interaction electroweak symmetry breaking.~\cite{PRM} 
Our aim is to explore whether this process allows us
to test experimentally the nature of the Higgs sector and the role
played by the top quark in the dynamics of the symmetry breaking.
This question has been addressed previously by Lee.~\cite{Lee}

The new strong interactions are expected to give the dominant 
contributions in scattering amplitudes with longitudinally 
polarized $W$ bosons. By the Equivalence Theorem, these are given
by Goldstone boson scattering amplitudes
$\pi^+ \pi^-\!\to t\bar t$.  
For  $\sqrt s >>m_t$, the low energy theorem (LET) gives
\bea 
&&{\cal M}( \pi^+ \pi^-\!\to t_R \bar t_L) =
- {2m_t^2 \over v^2} {\sin\theta \over 1 - \cos\theta},\hspace{1cm}
{\cal M}( \pi^+ \pi^-\!\to t_L \bar t_R) =  0, \nonumber\\
&&{\cal M}( \pi^+ \pi^-\!\to t_L \bar t_L) = {\cal M}(
\pi^+ \pi^-\!\to t_R \bar t_R ) = - {m_t \sqrt{s}\over v^2}.
\label{Eq:M}
\eea 
The helicity-flip amplitudes violate the $I=0$ partial wave 
unitarity bound at $\sqrt{s}\approx10$~TeV.\cite{CFH} The new physics 
of the symmetry breaking must therefore appear
below this scale. We expect, as in vector boson scattering,
that the new resonances predicted in strong 
interaction models will play a role in the unitarization.
Depending on the model, these may be the same resonances that
contribute to $W^+W^-$ scattering or new ones related to the 
mechanism of top mass generation. 
In the first case, illustrated by the SM, 
the top quark and $W$ boson couplings to the scalar should be 
proportional to the masses that they acquire by the Higgs mechanism, 
and Eq.(\ref{Eq:M}) become 
\be   
{\cal M}( \pi^+ \pi^-\!\to t_L \bar t_L) = {\cal M}(
\pi^+ \pi^-\!\to t_R \bar t_R ) =  {m_t \sqrt{s}\over v^2}
{M_S^2 \over s - M_S^2}\, .
\ee
The second possibility is illustrated by topcolor-assisted
technicolor models. These models have an additional scalar
(the top-Higgs) with an enhanced coupling to $t \bar t$ and a
suppressed coupling to $WW$. While the top-Higgs contribution to $WW$
scattering is negligible, its effect in $W^+ W^- \to t
\bar t$ is comparable to that of a SM Higgs boson of the same mass.
  
Some models of strong symmetry breaking, like technicolor, can also give
contributions coming from the s-channel exchange of  vector
resonances. The extended technicolor interactions that give rise to the
top mass also generate an effective coupling of the techni-rho ($\rho_T$)
to the $t \bar t$ pairs. The $\rho_T$ exchange contributes to the
helicity-conserving amplitudes with 
\be 
{\cal M}(\pi^+ \pi^-\!\to t_L\bar t_R, t_R \bar t_L) = 
- \eta_{L,R} {m_t \over v} {m_\rho
  \sqrt{N/2} \over v} {s \over s - m_\rho^2} \sin\theta,
\label{Eq:MTR}
\ee 
where N is the number of technifermion doublets and $\eta_{L,R}$ are
model-dependent dimensionless parameters of order one.  This
helicity-preserving contribution from a $\rho_T$ resonance is
comparable to the helicity-violating contribution from the scalar
exchange and can be the most important contribution in technicolor
models. Given this sensitivity to models, we investigate their
experimental signals in the next two sections.
 
\section{Signals in $e^+ e^-\!\to \nu \bar\nu t \bar t$}
\label{sec:2}

\begin{figure}[t]
\begin{center}
\vspace{-6cm}
\leavevmode
{\epsfxsize=11.8cm \epsfbox{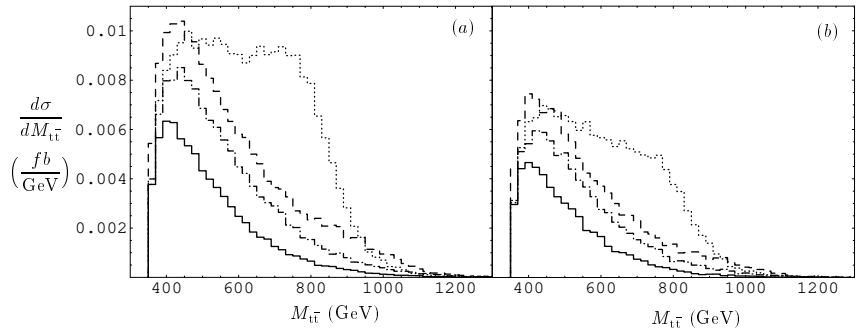} \vspace{-5.6cm}}
\end{center}
\caption{(a) Differential cross section for 
  $e^+ e^-\!\to \nu \bar\nu t \bar t$ as a function of the $t \bar t$
  invariant mass, for a LC with $\sqrt{s} = 1.5$ TeV. The curves
  correspond to the SM with a Higgs mass of 100 GeV (solid) and 800 GeV
  (dots), the technicolor model described in the text with $m_\rho$ =
  1 TeV (dash) and the LET limit (dot-dash).  (b) Same as (a)
  including ISR and beamstrahlung. \label{fig:1}}
\vspace{-2mm}
\end{figure}

The production of $t \bar t$ pairs from $W^+ W^-$ fusion can be
studied in a high energy $e^+ e^-$ LC by analyzing the reaction $e^+
e^-\!\to \nu \bar\nu t \bar t$.  We consider only fusion diagrams,
which are expected to dominate the production of $t \bar t$ pairs with
high invariant mass.  We use the effective-W approximation with
helicity-dependent structure functions~\cite{KA,PRM} of the $W$
bosons.  As we have seen in Sec.\ref{sec:1}, at extreme energies these
cross sections have a strong dependence on polarization, so we have
taken care to keep the helicity dependence in every stage of the
calculation. We use full helicity amplitudes for the process $W^+
W^-\!\to t \bar t$ and polarized top decay amplitudes to build up
polarization observables.

In our analysis, we have neglected the transverse momentum of the
$W$'s, expected to be of order $M_W$. Although this is a reasonable
approximation for computing the signal cross sections, the system
of cuts which separate the  $ W^+ W^-\!\to t \bar t$ events from 
the important backgrounds due to $\gamma \gamma \to t \bar t$ and 
$e^+ e^-\!\to t \bar t$ involves the total transverse momenta of the 
$t \bar t$ system.~\cite{SMR}  Thus, we do not address the 
backgrounds here.

In Figure \ref{fig:1}(a), we show the $t \bar t$ invariant mass
distributions in $e^+ e^-\to~\nu\bar\nu t\bar t$ at $\sqrt{s} =
1.5$ TeV predicted by the models described in Sec.\ref{sec:1}.  The SM
distribution with a Higgs boson mass of 800 GeV is compared with a Higgs
boson of 100 GeV, and the LET predictions. We have also
chosen a technicolor model with $N_{\rm TC}=4$ and  three
effective techniquark doublets. The $\rho_T$ mass is 1 TeV 
with a width of 45 GeV. The $\eta_L,\eta_R$ satisfy
the bounds imposed by $R_b$ measurements. In Figure
\ref{fig:1}(b) we show the effect of including initial state radiation
(ISR) and beamstrahlung (BS), using a 1.5 TeV NLC parameter 
set.~\cite{PE,AP}  Beamstrahlung is
responsible for roughly 80\% of the damping of the signal,
being more important at high values of $M_{t \bar t}$.  This
is precisely the region of interest for strong-interaction physics, 
and so it is gratifying that a
clear signal over the SM prediction with a Higgs of 100 GeV can still
be seen in all cases.

\begin{table}[bth]
\vspace{-0.4cm}
\caption{Number of $e^+ e^- \to \nu \bar\nu t \bar t$
events in a LC with $\sqrt s = 1.5$ TeV  and ${\cal L} = 200 fb^{-1}$,
for the models described in the text.
Event numbers including ISR, beamstrahlung, and cuts are also shown
separately. The estimated significance of the signal is given in brackets.
\label{tab:1}}
\vspace{0.4cm}
\begin{center}
\begin{tabular}{|l|rrrrrr|}
\hline
& &{\rm w.} && {\rm w.} &{\rm w. Cuts}& \\
Model& {\rm Events} &{\rm ISR+B} &&
{\rm Cuts} &{\rm ISR+B}& \\ \hline
{\rm SM}, $m_H$ = 100 GeV & 300 & 200 &(---)& 35 & 22 & (---) \\
{\rm SM}, $m_H$ = 800 GeV & 934 & 571 &(26)& 449 & 257& (50) \\
{\rm TC}, $m_\rho$ = 1 TeV & 661 & 419 &(15)& 233 & 133& (23)\\
{\rm LET} & 487 & 316  &( 8)& 139 & 82 &(13)\\ \hline
{\rm SM}, $m_H$ = 400 GeV & 1538 & 1092 &(63) &  & & \\ \hline
{\rm Top-C}, $m_{H_{t}}$ = 400 GeV & 974 & 668 & (33)&  & & \\ \hline
\end{tabular}
\end{center}
\end{table}

In Table \ref{tab:1}, we present the event yields for an
integrated luminosity of 200 $fb^{-1}$ and unpolarized beams, first
without and then with effects of ISR and beamstrahlung.
In models with heavy resonances, we have also
considered the effect of applying the cuts ($M_{t \bar t} > 500$ GeV,
$\cos\theta_{_{\rm CM}} < 0.5$ ) to enhance 
the strong-interaction signal.  In the second part of Table \ref{tab:1}, 
we compare the signal of a topcolor-assisted technicolor
model with $f_t/f_T = 1/4$ and a top-Higgs mass of 400 GeV, with the
signal of a 400 GeV SM Higgs boson.  Both models give an
increase of events at the same value of $M_{t \bar t}$ over the
light-Higgs expectations, but the top-Higgs signal is clearly smaller.
Finally, we give in parentheses an estimate of the significance of the
excess of events expected in each strong-interaction model over the
number of events expected in the SM with a Higgs of 100 GeV, which is
taken as background.  Although this is a highly idealized estimate of
the real significance of the signal, these numbers are 
comparable to those found by Barklow using a complete SM analysis.

\section{Top quark polarization observables}\label{sec:3}

Since scalar resonances couple to helicity-flip amplitudes and vector
resonances to helicity-conserving ones, a scalar resonance produces a
strong anti-correlation in the helicities of the final $t$ and $\bar
t$ quarks while vector resonance exchange tends to correlate the $t$
and $\bar t$ helicities.
Define  the helicity correlation asymmetry~\cite{SW} $C$ of the 
final $t \bar t$ pairs produced in $e^+ e^-\!\to \nu \bar \nu t \bar t$:
\be 
C = {\left[ \sigma (t_L \bar t_L) + \sigma (t_R \bar t_R) - 
\sigma (t_L \bar t_R) - \sigma (t_R \bar t_L) \right] \; /  
    \sigma({\rm unpolarized})} .
\label{C}
\ee
We find that this quantity has a strong dependence on the type of model.
 In the SM with M$_H$=100~GeV,
$t \bar t$ pairs are predominantly produced in the $RL$ configuration 
and $C = - 0.42$. 
If the Higgs mass is increased over the $t \bar t$ 
threshold, the number of $LL$ and $RR$ pairs is 
dramatically enhanced and $C$ takes a positive value.
For a Higgs mass of 800 GeV we find $C = + 0.51$. In technicolor models,
the combined effects of saturation in the $J=0$ channel and techni-rho 
exchange in the $J=1$  channel compete to give a small value of $C$. 
In the model described in Section \ref{sec:2} we find $C = - 0.12$. 
These values of C correspond to a 1.5 TeV LC and include ISR and 
beamstrahlung effects.

However, to extract this spin correlation asymmetry from the expected
sample of events will not be easy.  Since the $t \bar t$ system is
produced with smaller $M_{t \bar t}$ than the collider energy and a
substantial transverse momenta, we have to rely on fully reconstructed
events.  We will use for this analysis 6-jet and 4-jets+lepton 
decay modes, assuming an overall reconstruction efficiency (including
b-tagging) of $\epsilon_{\rm rec}=30$\%.
We assume that leptons can be identified with efficiency
$\epsilon_{\rm l}=100$\%.  Jet flavor
identification is needed in hadronic decays; to obtain this information,
we assume that it will
be possible to tag $c$ quarks with efficiency $\epsilon_{\rm c-tag}$=
50\%.

To get a first indication on the significance with which the spin of a
resonance could be probed, we work out the case of the SM with a Higgs
mass of 800 GeV.  To define observables correlated with the top quark
helicity, we use the angular distributions of the top quark decay
products 
\be 
{1 \over \Gamma} {d \Gamma_{t_{R,L}} \over d \cos \chi_i}
= { 1 \ \pm a_i \,\cos \chi_i \over 2},
\label{Eq:TDD}
\ee
where $\chi_i$ is the angle of the decay product $i$ with the top
helicity axis, measured in the top rest frame. For the $W^+$ decay
products, $a_l=1$ for $l=(l^+, d, s)$ and $a_\nu=-0.31$ for $\nu=(\nu,
u, c)$.  For the b quark, $a_b = -0.41$. The signs of $a_i$ are
reversed for top anti-quarks.  For each pair $(i,j)$ of $(t, \bar t)$
decay products and a particular model, we define $N_{ij}(+-)$ to be
the excess in the number of events with ($cos \chi_i >0$ , $\cos \bar
\chi_j < 0$) over the SM expectations with a Higgs of 100 GeV, and
similarly for the other sign combinations. Then, a scalar/vector
asymmetry can be defined for different types of $(t, \bar t)$ decay
products 
\be 
A_{ij}(S/V) = {1 \over a_i \bar a_j} { N_{ij}(-+) +
  N_{ij}(+-) - N_{ij}(++) - N_{ij}(--) \over N_{ij}(\rm{total})}
\label{Aij} 
\ee 
for a particular model and $A_{ij}(S/V) = 0$ by construction 
in the SM with a Higgs mass of 100 GeV. We will
use the decay angle of leptons and s-quarks (identified by a $c$-tag), which
have the stronger spin analyzing power $(a_l=1)$. The b quark can also
be used with a cut in the $W$ production angle $|\cos \chi_W
|<0.5$, which increases $a_b$ to $-0.6$ but with an additional cost in
efficiency $\epsilon_{\rm b-cut}$=50\%.  
For ${\cal L} = 200 fb^{-1}$ including ISR and BS, data samples 
corresponding to three different values of the $a_i$ should provide the
asymmetry  measurements:\\

\vspace*{-1cm}

$$
\begin{array}{llll}
A_{ll} (S/V) = 0.25 \pm 0.19 :  & (b  l \nu)(b  s c)  &
BR = 0.22,  & \epsilon = \epsilon_{\rm rec} \cdot \epsilon_{\rm l}
\cdot \epsilon_{\rm c-tag} \\
& (b   s c)(b  s c) &
BR = 0.11,  & \epsilon = \epsilon_{\rm rec} 
\cdot \epsilon_{\rm c-tag}^2 \\
A_{lb} (S/V) = 0.25 \pm 0.24 : & (b   l \nu)( b q \bar q)  &
BR = 0.44,  & \epsilon = \epsilon_{\rm rec} \cdot \epsilon_{\rm l}
\cdot \epsilon_{\rm b-cut} \\
 & (b s c)( b q \bar q)  &
BR = 0.22,  & \epsilon = \epsilon_{\rm rec} \cdot \epsilon_{\rm c-tag}
\cdot \epsilon_{\rm b-cut} \\
A_{bb} (S/V) = 0.25 \pm 0.63 : & ( b q \bar q)( b q \bar q )  &
BR = 0.44, & \epsilon = \epsilon_{\rm rec} 
\cdot \epsilon_{\rm b-cut}^2 
\end{array} 
$$
which combined give an
overall scalar/vector asymmetry $A (S/V) = 0.25 \pm
0.14$, a 1.8-$\sigma$ effect. With a 100\% left-handed polarized $e^-$ beam,
$A (S/V) = 0.25 \pm 0.10$, giving a statistical significance of the
scalar nature of the resonance of approximately 2.5-$\sigma$.

\section{Conclusions}

If a light Higgs boson is not found at present collider experiments,
it will be interesting to study the reaction 
$e^+ e^- \to t \bar t \nu \bar \nu $ at a high energy 
$e^+ e^-$ LC. We have found good signals in 
the $M_{t \bar t}$ distributions for strong-interaction Higgs sectors 
with scalar/vector resonances in the TeV region.
It may also be possible to probe the spin of these resonances by top 
quark polarization analysis.

\section*{Acknowledgments}

We are grateful to Tim Barklow for sparkling our interest in $t \bar
t$ production at high energy. ERM thanks the SLAC theory group for
their hospitality and J. F. de Troconiz for discussions.
The work of MEP has been supported by 
the US Department of Energy under contract DE-AC03-76SF00515.
ERM acknowledges support from the Spanish MEC and FAME under 
project AEN97-1678.

\end{document}